\documentclass[12pt,onecolumn,showpacs,amsmath,amssymb,aps,nofootinbib,floatfix]{revtex4-2}
\usepackage{graphicx} 
\usepackage{amsmath}
\usepackage{indentfirst} 
\usepackage{appendix}
\usepackage[top=3.0cm,left=3.0cm,right=2.0cm,bottom=2.0cm]{geometry}
\usepackage[utf8]{inputenc}
\newcommand{\be}{\begin{eqnarray}}

\newcommand{\lnz}{\ln \mathcal{Z}}

\newcommand{\ee}{\end{eqnarray}}

\newcommand{\ave}[1]{\left\langle #1 \right\rangle}
 \newcommand{\eqn}[1]{Eq.\,(\ref{#1})}

\usepackage{amssymb}
\usepackage[T1]{fontenc}
\usepackage{cleveref}
\usepackage{titlesec}
\usepackage{xcolor}
\titleformat{\chapter}
  {\normalfont\fontsize{14}{14}\bfseries}{\thechapter}{1em}{}
\titleformat{\section}
  {\normalfont\fontsize{14}{14}\bfseries}{\thesection}{1em}{}
\titleformat{\subsection}
  {\normalfont\fontsize{14}{14}\bfseries}{\thesubsection}{1em}{}
\titleformat{\subsubsection}{\normalfont\fontsize{14}{14}\bfseries}{\thesubsubsection}{1em}{}

\begin{document}
\title{Gaussian non relativistic spontaneously stochastic hydrodynamics}
\author{David Montenegro, Giorgio Torrieri}
\affiliation{Universidade Estadual de Campinas - Instituto de Fisica Gleb Wataghin\\
Rua Sérgio Buarque de Holanda, 777\\
 CEP 13083-859 - Campinas SP\\
}
\begin{abstract}
We study the non-relativistic limit of Gaussian covariant hydrodynamics [1]. We argue that the condition of incompressibility provides additional symmetries matching relativistic hydrodynamics but incompressibility must break down at a ``microscopic`` scale. We then develop the renormalization group equations for average and fluctuations w.r.t. that scale, to understand its effect on flows at intermolecular distances where hydrodynamics gives way to statistical mechanics. The resulting dynamics naturally incorporates spontaneous stochasticity as a macroscopic back reaction of statistical mechanics fluctuations, as well as features reminiscent of anomalous dissipation and ``wild solutions`` as renormalization group counterterms. We frame these considerations into both a phenomenological discussion of the limits of applicability of fluid dynamics, and a discussion of where physics might shed some light on the mathematical issues associated with turbulence.
\end{abstract}
\maketitle
\section{Introduction}
\subsection{The background issues}
The dynamics of fluids, while traditionally a domain of engineering departments (as ``hydraulics``, cynically labelled ''observing phenomena which could not be explained'') or of mathematics departments (as ``fluid mechanics``, equally cynically  labelled '' explaining phenomena which could not be observed'') \cite{dalam} is a profoundly physical phenomenon.

Physicists tend to think of fluids as described by an effective theory written in terms of a gradient expansions of conserved currents of a continuum which is locally close to thermal equilibrium \cite{delacretaz}.   The frame in which this equilibrium is defined is called the hydrodynamic flow frame \cite{kambe}, and it is also the frame in which scalar conserved currents are at rest.    The equilibrium parameters (density, temperature and equation of state) together with the flow then determine the dynamics via conservation laws via a series in gradients weighted by a small dimensionless parameter, the dissipative over macroscopic gradient scale,usually known as the Knudsen number.   In this sense, therefore, hydrodynamics is a traditional effective field theory.
The great mysteries of hydrodynamics, such as the existence of solutions of the Navier Stokes equations, are usually thought of as abstract mathematics \cite{millenium,hilbert} in which physics has little or no role, even when the questions asked, such as anomalous dissipation \cite{andiss} and the existence of Wild/Nightmare solutions \cite{wild} appear physically relevant.

However, looking carefully there are mysteries within the physical aspects of fluid dynamics which might challenge this consensus.  Away from perfect equilibrium, the flow frame is not so well defined because of diffusive currents, leading to several definitions of flow frames \cite{disconzi,rocha} (in non-relativistic systems the  frame at rest with a conserved charge, i.e. the mass density, is usually far more convenient, but the conceptual problem does remain).

This ambiguity masks a deeper problem:  Hydrodynamics depends on statistical mechanics, for the equation of state and eventually transport coefficients.
But ``Equilibrium`` in statistical mechanics \cite{huang} means that all microstates are equally likely, and therefore the macrostate with most microstates is overwhelmingly probable.   However, there is no real definition of what approximate equilibrium within a slowly changing environment in this microstates language \cite{jaynes}.   More concretely, there is no systematic theory combining the dynamics of stochastic fluctuations with the gradient expansion.    Attempts for such a theory have been made with linear response \cite{tong,forster,kadanoff} and Feynman diagram techniques \cite{kadanoff,kovtun}.  However in the relativistic regime such approaches tend to conclude that hydrodynamics is extremely non-universal \cite{jain} away from the continuum limit, in contrast with experiment at both very high (relativistic) \cite{zajc},  very low  energy \cite{giacalone} energies and even everyday situations (the ''Brazil nut effect'' \cite{braznut})

To appreciate just how potentially revolutionary are the results described in \cite{zajc,giacalone}, one should look at the supposed resolution of Hilberts 6th problem which was rewarded with a FIelds medal in 2026 \cite{hilbert}.
But the proof relies crucially on the Grad limit, the assumption that the number of degrees of freedom diverges parametrically faster than the inverse of the Knudsen number.  This is exactly what results such as \cite{zajc,giacalone} show might not be applicable to ''real'' fluids.  Thus, the linked problems of ''what is the smallest fluid?'' and ''how do thermal fluctuations backreact on fluid evolution?'' are quintessentially physics questions, to be studied via experimental and theoretical tools familiar from physics.

In the non-relativistic limit, the phenomenon of spontaneous stochasticity \cite{spont1} suggests that fluctuations are not just perturbations on top of a deterministic theory but emerge in the same regime as where a fluid emerges. It should be noted, in this regard, that spontaneous stochasticity has been shown to allow molecular fluctuations, i.e. exactly the fluctuations determined by microstate distributions \cite{jaynes} to communicate with ``mesoscopic`` turbulent scales \cite{spont2}.  This seems to provide deep insights into mathematics issues such as Onsager's conjecture and anomalous dissipation (which becomes driven by statistical fluctuations), but the scale separation between the microscopic and mesoscopic scale is puzzling.

Because the issue is the effect of fluctuations on scale separation, 
quantum field theory techniques such as renormalization \cite{peskin} therefore look like a promising tool \cite{onsager,hnatic,calzetta,mccomb} but the seemingly non-perturbative nature of fluid dynamics in such a regime makes a quantitative calculation problematic, as non-perturbative renormalization group techniques \cite{frg1,frg2} are generally inapplicable to strongly time-varying problems

Recently, a rederivation of hydrodynamics has been proposed that explicitly deals with these issues \cite{gaussian,gaussiandiff,crooks,ergodic,gavassino}.  The idea is that rather than treat the fluid as a deterministic system evolving through conservation equations (with gradient-expanded constitutive relations giving conserved currents), it can be thought of as a Wiener/Ito process\cite{ito}, determined by an evolving partition function whose first two derivatives are determined non-perturbatively and constrained by Ward identities. To leading order \cite{gaussian} there is a partition function in every cell but it is always of Gaussian form. This approach has two advantages:  Obviously, fluctuations can  be included from the start.
But also, the existence of a partition function can be used to treat the symmetries of fluid dynamics via Ward identities.   Thus, the symmetries of ideal fluid mechanics are linked to microscopic ergodicity of the microscopic degrees of freedom making up the fluid \cite{ergodic}.   Then, the ambiguity of the definition flow in the relativistic regime \cite{disconzi,rocha} becomes a consequence of the general covariance of the theory.   The purpose of this work is to explore the non-relativistic limit of such a picture, with a view to make a link with both spontaneous stochasticity and the local symmetries of non-relativistic fluid dynamics \cite{fluidsym}, and eventually the more mathematical issues discussed in the introduction. 

First of all, one must be careful about what a non-relativistic limit means \cite{jensen}.  The Galileo group is not a subgroup of the Lorentz group so one can not get a unique limit by taking a certain number of generators to zero.
Physically, a non-relativistic continuum is obtained by taking $c\rightarrow \infty$ and, concurrently, $\mu/T \rightarrow \infty$ so that the particle density is kept finite.   But there is no unique way to do that and it generally results in a very different gradient expansion w.r.t. the relativistic one \cite{liaokoch}.

Symmetries provide a roadmap on how to proceed:  It is easy to note that volume preserving diffeomorphism invariance is the symmetry of ideal relativistic hydrodynamics \cite{dubovsky} and also of ideal {\em incompressible} non-relativistic hydrodynamics \cite{arnold}. Physically, volume preserving diffeomorphisms ensure the existence \cite{dubovsky} of a conserved entropy current and a Killing vector (flow). As argued in \cite{gaussian}, if one interprets this entropy current in a Gibbsian way, it is reasonable that volume preserving diffeomorphisms survive in a fluctuating regime because locally entropy creation can not be distinguished from a thermodynamic fluctuation.

Thus, incompressibility concurrently with the non-relativistic limit, i.e. taking the speed of sound $c_s$ to infinity together with the speed of light might give a consistent limit. 
It is empirically known that non-relativistic fluids tend to be incompressible in the low viscosity limit, especially in the turbulence regime (where incompressibility is related to scale-invariance \cite{kolscale}). The sort of strong interactions that ensure low viscosity will also, given the Pauli exclusion principle, ensure a large speed of sound.   Of course too strong interactions in the non-relativistic regime can drive a fluid into a solid phase, where there is no volume preserving diffeomorphism symmetry and also no local equilibrium \cite{ergodic} (solids are brittle).
However, in addition to relativity, incompressibility is incompatible with an analytical statistical partition function, something obvious from the non-local definition of pressure \cite{ghallagher1,ghallagher2} as well as explicit computations from the equation of state \cite{pathria,cswalecka,sorensen}.

The incompressibility assumption also invalidates the Boltzmann equation (from which  hydrodynamics is often derived \cite{hilbert,rocha}), as it is based on sequential scattering, typically leading to ideal gas type laws which have a finite speed of sound.
So far in the literature the problem has been essentially treated by combining the Boltzmann equation with the assumptions that Strouhal and Mach numbers are negligible on the same scale as the Knudsen number \cite{golse}.   This contradicts the actual non-relativistic kinetic equation estimate \cite{cerci}, where $\lim_{m \rightarrow \infty} c_s^2 \sim (T/m)^{1/2} \rightarrow 0$   In summary, the literature on the topic is somewhat inconsistent:  Incompressibility is taken to be ''fundamental'' in the mathematical treatment of non-relativistic hydrodynamics, but everybody knows it is at best an effective description.  What makes this troubling is that it is not clear whether the incompressibility approximation is relevant at regimes where both turbulence and spontaneous stochasticity play a role, a question crucial in the relativistic limit \cite{eyinkspont}.

Finally, the flow ambiguity that plays such a central role in \cite{gaussian} is collapsed by incompressibility:  The Bernoulli equation \cite{kambe}, in its exact form $\nabla.u=0$ renders the definition of flow unique. Usually the speed of sound is handled via a relation to the equation of state. From a general partition function and the Gibbs-Duhem relations \cite{pathria,sorensen,cswalecka}
However, as argued in \cite{gaussian}, a thermostatic equation of state breaks general covariance;  A generally covariant hydrodynamics will have fluctuations and averages evolving in a way constrained  by the gravitational Ward identity.
Together with the linear response equation, this exactly constrains the parameters of a Gaussian partition function provided this function is Gaussian.    

A recent topological demonstration that incompressibility uniquely determines fluctuations and their relation with dissipation \cite{braunstein2} , via the decoupling to all orders of the microscopic and convective backreactions, offers evidence that these issues are resolvable within functional renormalization group techniques \cite{frg1,frg2} since the Gaussian functional is a renormalization group fixed point \cite{iona}.   In this picture, at microscopic scales the fluid is indeed compressible, but this manifests itself at scales unobservable in experiments made with fluids (in practice, to probe intermolecular distances fluids are inevitably destroyed).  However, if one looks at fluctuations and their effect on evolution, remnants from the compressible scale manifest themselves in counterterms, which accommodate anomalous dissipation dynamics.

The next subsection will give the basic structure of how \cite{gaussian} and \cite{gaussiandiff} play out in the non-relativistic limit.   In the next sections, the details of the evolution, including non-relativistic Ward identities \cite{wardgal,geracie,brauner} and functional renormalization group evolution \cite{frg1,frg2} combine with this picture.
\subsection{The basic setup}
Usually hydrodynamics \cite{kambe} is written in terms of energy density $e$, pressure $p$, molecular density $n$ as well as a flow vector $u_i$ (or equivalently $\beta_i=u_i/T)$,from which one builds, via constitutive relations, the energy momentum conserved current $T_{ij}$ and the molecular current $J_i$.  

The basic idea behind 
 \cite{gaussian,gaussiandiff}, is to try to work directly with the conserved currents and their non-perturbative fluctuations by working out the evolution of the partition function, but approximating it at all times with a Gaussian
\begin{equation}\label{partition}
\mathcal{Z}=\exp\left[- \int_\Sigma \left( \beta_i T^{ij}+\mu J_j \right) d\Sigma_j \right] \simeq  \prod_{x,x'}
\exp \left[ -\frac12
\begin{pmatrix}
J-\bar J\\
T-\bar T
\end{pmatrix}^{\!T}_x
\begin{pmatrix}
D & M^{T}\\
M & C
\end{pmatrix}^{-1}_{x,x'}
\begin{pmatrix}
J-\bar J\\
T-\bar T
\end{pmatrix}_{x'}
\right].
\end{equation}
where $d^3\Sigma_i$ is an arbitrary vector with dimensions of the volume (the coordinates of the Lagrangian particle/volume cell),  $T\equiv T_{ij}-\ave{T_{ij}(x)},J\equiv J_i-\ave{J_i(x)}$ are the conserved currents and $C_{ijkl},M_{ijk},D_{ij}$ are the correlators between currents\footnote{note that the exponential is a scalar}.   Making $d^3\Sigma_i$ a vector with a velocity allows us to build lagrangian co-moving coordinates of a cell and, crucially, to implement the symmetries characterizing ideal hydrodynamics \cite{dubovsky,fluidsym}, also in a fluctuating environment (where these symmetries give rise to ghost-like redundancies \cite{gaussian,gavassino}) and in the distribution of microstates \cite{ergodic}.

The basic idea in \cite{gaussian} is that constraints from Ward identities and linear responses giving equations of motion for $\ave{T_{ij}(x)},\ave{J_{i}(x)}, D_{ik}(x,x'), M_{ijk}(x,x'), C_{ijkl}(x,x')$.   Thus, the initial condition is not a field of flows, densities and pressures, but rather an ensemble of currents characterized by $\ave{T_{ij}},\ave{J_{i}}, D_{ik}, M_{ijk}, C_{ijkl}$.
Once this ensemble is given at a given time, however, it can be propagated forward, thus ensuring a non-perturbative stochasticity and every timestep and the observation of local symmetries, which is related to strong hyperbolicity \cite{gaussiandiff}.

The  Ward identity for Galilean invariance, a  Noether  second theorem relation relating the first and second cumulant 
\begin{equation}
\label{genward}
\dot{\mathcal{L}}_{0 \ldots }-\nabla.\mathcal{L}
_{ij \ldots}=0, \qquad \mathcal{L}_{ij...}=\mathcal{O}_{[2] ij}-\delta(x_1-x_2)\sum_j a_j \mathcal{O}_{[1]j}, \qquad \mathcal{O}_{[n]} \equiv \frac{\delta^n \lnz}{\delta \mathcal{J}^n}
\end{equation}
where $\mathcal{O}_1$ can refer to averages,   $\ave{T_{ij}},\ave{J_{i}}$, and $\mathcal{O}_2$ to second cumulants $D_{ik}, M_{ijk}, C_{ijkl}$ 
(eq. 12 of \cite{gaussian} gives a concrete example. Here $\mathcal{J}$ is a generic source).

By the linear response from \cite{tong,kadanoff,forster}, we mean an integral equation obtained from an analytical continuation of $\mathcal{O}_2$
\begin{equation}
\mathcal{O}_{[1]}(t+\Delta t)=\int_t^{t+\Delta t} dt^\prime  \mathcal{O}_{[2]}(t,t^\prime ) \mathcal{O}_{[1]}(t^\prime ), \qquad \tilde{\mathcal{O}}(t,k)=\frac{1}{2i}\left(\frac{\tilde{\mathcal{O}}(t,k)}{\tilde{\mathcal{O}}(-i\epsilon t,k)}-1\right)
\label{linrespdef}
\end{equation}
\eqn{genward} and \eqn{linrespdef} are two equations with two unknowns, $\mathcal{O}_{[1.2]}$ so in principle they are exactly the ingredients needed to propagate the parameters of \eqn{partition} from their initial condition values.

Non-perturbative Renormalization group flows can then be implemented  at the operator lever by considering the Fourier transforms of the operators.
\begin{equation}
\label{renfourier}
\tilde{\mathcal{O}}_{1,2 i_1,i_2,...}(\vec{k},t)=\int  d^3 x \, \mathcal{O}_{1,2 i_1,i_2,...} \exp\left [i\left(\vec{k} \cdot \vec{x}\right)\right]=\left\{ \begin{array}{cc}
\tilde{\mathcal{O}}^{UV}_{1,2} & k\geq k_0\\
\tilde{\mathcal{O}}^{IR}_{1,2} & g \leq k_0
\end{array} \right. 
\end{equation}
where IR/UV distinguishes between the infrared (incompressible) and ultraviolet (compressible, and eventually fully relativistic) regimes. The Gaussian approximation implicit in \eqn{partition} allows us to determine a renormalization group flow \cite{peskin,frg1,frg2}.
\begin{equation}
\lnz \left[ k_0,J,T,D,M,T\right]=\lnz\left[ k_0+\Delta k_0,J+\Delta J,T+\Delta T,D+\Delta D,M+\Delta M\right]
\end{equation}
and to find the fixed point of this evolution. In the Gaussian Wiener/Ito process, this is all we need.

In the rest of the paper, we shall fill in these ingredients in detail.   We start with finding the second order Galilean group Ward identity, using the methods of  \cite{wardgal,geracie,brauner}, for both compressible and incompressible fluids.  We then work out in detail the renormalization group equations relating the two, and find the exact form of the linear response equations.

\section{The evolution of the fluctuation tensor: Ward identities and the renormalization group flow}
\subsection{Ward identities}
We start with computing the second order Ward identities in the Galilean case, following the Newton-Cartan formalism introduced in \cite{wardgal,geracie} and proceeding in Fourier space, where $\partial_i \rightarrow k_i,\frac{d}{dt}\rightarrow w$. We have for the fundamental connected correlators $ D_{ik}(x,y) = \left\langle J_i(x)J_k(y) \right\rangle_c, \ M_{ijk}(x,y) = \left\langle T_{ij}(x)J_k(y) \right\rangle_c, \ C_{ijkl}(x,y) = \left\langle T_{ij}(x)T_{kl}(y) \right\rangle_c $ decomposing the current and stress tensor as $ J_i=J_i^{\mathrm{macro}}+j_i^{\mathrm{micro}},
\ T_{ij}=T_{ij}^{\mathrm{macro}} + \tau_{ij}^{\mathrm{micro}} $. The Ward identities must therefore take the form
\begin{align}\label{cac1}
\omega D_{ik}^{\mathrm{macro}} - k_j M_{ijk}^{\mathrm{macro}} - \Delta_{ik}^{JJ}=0,
\\ \label{cac2} \omega M_{kli}^{\mathrm{macro}} - k_j C_{ijkl}^{\mathrm{macro}} - \Delta_{ikl}^{JT}=0,
\end{align}
 where $ \Delta_{ik}^{JJ} = - \omega D_{ik}^{\mathrm{mixed}} + k_j M_{ijk}^{\mathrm{mixed}}, \Delta_{ikl}^{JT} 
= - \omega M_{kli}^{\mathrm{mixed}} + k_j C_{ijkl}^{\mathrm{mixed}}$ \footnote{The mixed correlation function is $ \left\langle \mathcal{O}^{\mathrm{macro}}, \mathcal{O}^{\mathrm{micro}} \right\rangle$}. The corrections satisfy constraints coming from Galilean symmetry. Putting everything together in the \eqref{cac1} and \eqref{cac2}, the Ward identities will have the form for the incompressible construction ($k_0=\omega$ is also used in the non-relativistic equation for simplicity)

\begin{equation}
\omega^{2} ( D^{\mathrm{macro}}_{ij} - D^{\mathrm{mixed}}_{ij} ) - 2 \omega k^l M^{\mathrm{mixed}}_{ijl}  - k^l k^k ( C^{\mathrm{macro}}_{ijlk} - C^{\mathrm{mixed}}_{ijlk} ) = 0
\begin{cases}
\mathcal{O}^{\mathrm{macro}},
& |k|\leq k_{0},\  |\omega|\leq k_{0}/c,
\\[0.6em]
\mathcal{O}^{\mathrm{mixed}},
& |k|\geq k_{0},\  |\omega|\leq k_{0}/c,
\\[0.6em]
\mathcal{O}^{\mathrm{micro}},
& |\omega|\geq k_{0}/c.
\end{cases}
\label{allward}
\end{equation}
i.e. $\mathcal{O}^{\mathrm{macro}}$ the non-relativistic incompressible form for $k\leq k_0,w\leq k_0/c$, $\mathcal{O}^{\mathrm{mixed}}$ non-relativistic compressible for $k\leq k_0,w\geq k_0/c$ and $\mathcal{O}^{\mathrm{micro}}$ relativistic otherwise.  At the critical frequencies the terms need to match, but not their derivatives, for reasons that we shall explain in the remainder of the paper. The correlations function become $D_{ik}^{\mathrm{macro}} = \rho_0^2 \left\langle u_i u_k \right\rangle_c,$ and $ k_j k_l C_{iljk}^{\mathrm{macro}} = -\eta^2 k^4 \left\langle u_i u_k \right\rangle $. For $k \leq k_0 \equiv \rho_0^{1/3}$ the fluid is effectively incompresible (so the only correlators are velocity and pressure) above that scale compressibility breaks down. At that point the Ward identity will be either the fully compressible one, or eventually the Lorentz covariant one derived in \cite{gaussian} and extended to finite chemical potential in \cite{gaussiandiff}.

The relativistic fluctuating statistical mechanics can, for the purposes of this paper, be considered UV-complete since a Gaussian renormalization group fixed point should exist \cite{gaussian,iona}.

\subsection{Renormalization group structure}
It seems physically inappropriate here to put the full Lorentz invariant Ward identity, given that we know that for most fluids (such as water) velocities at molecular distances are still well below relativistic effects, so it might be logical to concentrate on just the first two columns of \eqn{allward}, and this might well be the case. However, this is physically irrelevant, since our purpose will be to use the Ward identities and a Gaussian ansatz to build a functional renormalization group (FRG) \cite{peskin,frg1,frg2} type equation linking the incompressible non-relativistic and the relativistic limit (which are linked by the local volume preserving diffeomorphism symmetries). Physically, the point is that both the compressible and the Lorentz scale are inaccessible by experiments where the fluid is there, and the FRG implements this invisibility in the dynamics, with the residual effects of the macroscopic dynamics in the counterterms.

We speculate that corrections to the averages $\ave{J(x)},\ave{T(x)}$ (\eqn{partition}), represent what mathematicians call anomalous dissipation \cite{andiss}, roughly the backreaction of stochastic fluctuations at the molecular scale to macroscopic dissipative currents. The corrections to the fluctuations $C,M,D$ represent "wildness" (from what mathematicians call Wild solutions \cite{wild}), I.e. the spontaneous microscopic turbulence undetectable on macroscopic scales.

\textcolor{black}{We realize this is highly speculative for  concepts  such as wild solutions, and conjectures such as Onsager`s arise from abstract mathematics, often within highly idealized non-physical definitions used in rigorous proofs \cite{wild,onsager,andiss}. But we point out to the conceptual similarity of weak solutions, on which such results are based\cite{andiss} to what physicists call coarse graining, and universality under test function choice to what physicists mean by renormalization\cite{peskin}, to argue this conjecture is not implausible}

Using the notation \cite{frg1}, the standard Legendre transformations are
\begin{equation}
W_k[\mathcal{J},T] = \ln Z_k[\mathcal{J},T],
\end{equation}
Thus,
\begin{equation}
\mathcal{J}_i = \frac{\delta W_k}{\delta a_i},
\quad T_{ij} = \frac{\delta W_k}{\delta b_{ij}}.
\end{equation}
The Ward identities and the Gaussian ansatz are the tools used to construct and constrain this flow, but they are not separate endpoints or additional physical regimes.
\begin{equation}
\Gamma_k [\Phi] = \Phi \cdot \left(
\begin{smallmatrix}
J\\
T
\end{smallmatrix}
\right) - W_k[\mathcal{J},T] - \Delta S_k[ \Phi ],
\end{equation}
where $ \Delta S_k[\Phi] = \frac12 \int_q \Phi_A(-q)\, R_{k,AB}(q)\, \Phi_B(q)$. The exact flow equation becomes
\begin{equation}\label{123}
\partial_k\Gamma_k = \frac12
\operatorname{Tr} \left[ \left( \Gamma_k^{(2)} + R_k \right)^{-1} \partial_kR_k \right]
\end{equation}
where $ t=\ln\left(\frac{\Lambda}{k}\right)$ and the $ \partial_k R_k $ contains not only the ordinary flow terms, but also a contribution proportional to $ \delta(k-k_0) $, which represents the sharp matching at the compressibility threshold, indicated in \eqref{allward}, and acts as a momentum-shell selector.
\begin{equation}\label{reg}
\partial_k R_k = \Theta(k_0 -k) \, \partial_kR_k^{<} + \Theta(k-k_0) \, \partial_kR_k^{>} + \delta(k-k_0) \left( R_k^{>} - R_k^{<} \right)
\end{equation}
where $R_k^{<}$ and $R_k^{>}$ are the regulators used in the infrared incompressible and compressible ultraviolet theory, respectively. The last term in \eqref{reg} accounts for the abrupt change in the set of active degrees of freedom at the crossover scale. Low-momentum modes are suppressed by a scale-dependent quadratic term, producing a scale-dependent effective action that interpolates between microscopic and full effective descriptions. The full propagator is
\begin{equation}\label{greenf}
G_{\kappa,AB}(q)
= \bigg( \Gamma_k^{(2)} + R_k \bigg)^{-1} =
\begin{pmatrix}
D_{\kappa,ij}(q)
&
M_{\kappa,j\mid kl}(-q)
\\[6pt]
M_{\kappa,i\mid lm}(q)
&
C_{\kappa,ij\mid lm}(q)
\end{pmatrix}.
\end{equation}
where $\kappa$ is the running scale. As incompressibility is a macroscopic limit, not a fundamental microscopic identity.
\begin{equation}
\Gamma_k
\longrightarrow
\begin{cases}
\Gamma_{\mathrm{inc,NR}},
& \text{for } k \text{ in the incompressible non-relativistic regime},
\\[4pt]
\Gamma_{\mathrm{rel}},
& \text{for } k \text{ in the compressible relativistic regime}.
\end{cases}
\end{equation}
This is the complete formal FRG skeleton:
\begin{equation}
\Gamma_{\mathrm{rel}}
\;\xrightarrow[\;k:\,\Lambda\rightarrow0\;]{\text{Wetterich equation}}\;
\Gamma_{\mathrm{inc,NR}}.
\end{equation}
Write the effective action as
\begin{equation}\label{wett}
\Gamma_{\kappa} = \Gamma_{\kappa,\mathrm{inc}}
+ \Gamma_{\kappa,\mathrm{comp}}
+ \Gamma_{\kappa,E},
\end{equation}
where the sectors are selected in \eqref{allward} by the appropriate step functions and the boundary conditions are in \eqref{boundaryt}. Following the standard Wetterich construction, the infrared effective average action is obtained through the modified Legendre transform,
\begin{equation}
\Gamma_{\kappa}^{\mathrm{inc}}[\Phi_{>}]
= \Phi_{>} \cdot \left(
\begin{smallmatrix}
J\\
T
\end{smallmatrix}
\right) - W_{\kappa}^{\mathrm{inc}}[J_T] - \Delta S_{\kappa}[\Phi_{>}],
\label{eq:GammaInc}
\end{equation}
The effective average action can be decomposed as
\begin{equation}
\Gamma_{\kappa}^{\mathrm{comp}}[\Phi_{<}]
= \Gamma_{\mathrm{NS}}[\Phi_{>}] + \Delta\Gamma_{\kappa}[\Phi_{>},\Phi_{<}],
\label{eq:effective_inc}
\end{equation}
where \(\Gamma_{\mathrm{NS}}\) is the classical incompressible Navier--Stokes action, and \(\Delta\Gamma_{\kappa}\) contains renormalizations of the transport coefficients, nonlocal interactions, higher-order transverse vertices, noise corrections, and every operator compatible with the symmetries of the incompressible effective theory. Consequently, the infrared dynamics is entirely described on the transverse field manifold,
\begin{equation}
\Gamma_{\kappa\ll k_0}^{\mathrm{inc}} [\Phi_>] = \Gamma_{\kappa\ll k_0}[\Phi],
\label{eq:IR_inc}
\end{equation}
although the underlying microscopic theory remains fully compressible. The transverse one-particle-irreducible two-point function obtained from the compressible theory must reproduce the incompressible effective theory at momenta much smaller than the crossover scale,
\begin{equation}
\Gamma^{(2),\mathrm{comp}}_{TT}(\kappa)
= \Gamma^{(2),\mathrm{inc}}(\kappa)
+ \mathcal O \!\left[ \left( \frac{|\kappa|}{k_0} \right)^n \right], \qquad
|\kappa| \ll k_0,
\label{eq:matching_2point}
\end{equation}
where \(n>0\) is determined by the leading irrelevant operator generated during the elimination of the compressible sector\footnote{We present the matching condition in \eqref{boundaryt}.}. The terms suppressed by powers of $n$ correspond to the compressible microscopic theory and the incompressible effective theory described by the same low-energy physics. Equivalently, the transverse propagators satisfy
\begin{equation}
G_{eff|ij}^{\mathrm{comp}}(\kappa) =
G_{ij}^{\mathrm{inc}}(\kappa), \qquad
|\kappa| \ll k_0.
\end{equation}
where $G_{eff|ij}^{\mathrm{comp}}$ is the effective propagator after integrating out the compressible fields. These matching conditions guarantee that the low-energy transverse observables are independent of whether they are computed directly in the compressible microscopic theory or in the incompressible effective
theory, up to corrections suppressed by powers of \(|\kappa|/k_0\).

\subsection{Compressible Fluids}

The scale-dependent effective average action is written as
\begin{equation}
\Gamma_{\kappa} = 
\Gamma_{\kappa}^{\mathrm{continuity}}
+ \Gamma_{\kappa}^{\mathrm{momentum}}
+ \Gamma_{\kappa}^{\mathrm{noise}}.
\end{equation}

For the shear viscosity
\begin{equation}
\label{microshear}
\partial_t\eta_{\kappa} =
\rho_0 \left. \frac{\partial}{\partial p^2}
\, \partial_t \Gamma_{\kappa,\bar{\pi}_a\pi_a}^{(2)}
(\omega,\mathbf{p}) \right|_{\omega=0,\mathbf{p}=0}.
\end{equation}
Here $\pi_i = \rho u_i$ denotes the momentum density, and $\bar{\pi}_i $ is the corresponding response field\footnote{$\bar{\pi}$ is introduced to formulate the stochastic dynamics and generate response functions, rather than an independently measurable hydrodynamic variable, so that the stochastic equation of motion can be written as an action.}. For the sound speed
\begin{equation}
\label{runsound}
\partial_t c_{s,\kappa}^{\,2}
= \frac{1}{i} \left. \frac{\partial}{\partial p} \, \partial_t \Gamma_{\kappa,\bar{\pi}_1\varrho}^{(2)}
(0,\mathbf{p}) \right|_{\mathbf{p}=0}.    
\end{equation}
Here
$c_{s,\kappa}^{\,2}=
\left.\partial p_{\kappa}/\partial\rho\right|_{\rho_{0,\kappa}}$
is the running adiabatic pressure response, i.e the speed of sound. 

Usually the speed of sound is thought to be inherently related to the equation of state, via  \cite{sorensen} through the Gibbs-Duhem relations
\begin{equation}
\label{thermsound}
\ln\left( \frac{T}{T_0} \right) = \int_{\mu_0}^\mu \frac{ \rho' d\mu}{\frac{c_s^{2}}{\rho'}\left(\frac{ d \rho'}{d p}\right)_s -\mu \rho'}
\end{equation}
\eqn{runsound} can be thought of as a correction to this due to hydrodynamic response to thermal fluctuations looking like a compressible sound wave.  In the long-wavelength incompressible regime, this integral diverges so these perturbations would be indetectable at the bulk (they would correspond to ``fluctuations of the boundary``, by definition undetectable).   But a microscopic compressible scale means that \eqn{thermsound} will receive contributions from the counter-term \eqn{runsound}.

Similarly For the bulk (breathing mode) component,
\begin{equation}
\label{runbulk}
\partial_t\zeta_{\kappa} = \rho_0 \left. \frac{\partial}{\partial p^2} \, \partial_t \Gamma_{\kappa,\bar{\pi}_1\pi_1}^{(2)}
(\omega,\mathbf{p}) \right|_{\omega=0,\mathbf{p}=0}
- \frac{3}{4} \partial_t\eta_{\kappa}.
\end{equation}
Once again, an incompressible limit would make \eqn{runbulk} undetectable, but if incompressibility is cutoff at a critical scale this mode will appear. 
  As we argued earlier, for non-relativistic fluids these modes are themselves considered fluctuations, since the breakdown of compressibility (where such sound and bulk modes are possible) only appear on microscopic scales.
  Thus, these modes will be associated to counterterms to the fluctuation operator.

\subsection{Incompressible Fluids}

The incompressible effective average action is
\begin{equation}
\Gamma_{\kappa}[u,\bar{u}] = \Gamma_{\kappa,2} + \Gamma_{\kappa,3},
\end{equation}
The evaluation of the Wetterich trace in appendix
\begin{equation}\label{trace}
\partial_t\Gamma_{\kappa} = \frac{1}{2} \operatorname{Tr} \left[ G_{\kappa}\, \partial_tR_{\kappa} \right] = \frac{\nu_{\kappa}  \kappa^{5} }{3\pi Z_{\kappa}} \,.
\end{equation}
where $\nu_{\kappa}$ is the running kinematic viscosity, and $Z_{\kappa}$ is the wave-function renormalization ($ u_i = Z_{\kappa}^{-1/2} u_i^{(\mathrm{R})} $), $\kappa$ comes entirely from dimensional analysis.
\begin{align}
\left. \partial_t \Gamma^{(2)}_{\kappa,\bar{u}_i u_j}(p) \right|_{3\times3} &= \widetilde{\partial}_t \int_q V_{iab}(p,q)\, C^\mathrm{inc}_{\kappa,aa'}(q)\, G^{R}_{\kappa,bb'}(p-q)\, V_{ja'b'}(-p,q-p) \nonumber \\ 
&=  \widetilde{\partial}_t
\int_{\omega^\prime,\mathbf{q}} \frac{ 2D_{\kappa}(\mathbf{q}) }{ Z_{\kappa}^{2} \omega^{\prime 2} + A_{\kappa}^{2}(\mathbf{q})} \, \frac{ V_{iab}(\mathbf{p},\mathbf{q})\, P^{T}_{aa'}(\mathbf{q})\, P^{T}_{bb'}(\mathbf{p}-\mathbf{q})\, V_{ja'b'}(-\mathbf{p},\mathbf{q}-\mathbf{p}) }{ -iZ_{\kappa}(\omega-\omega^\prime) + A_{\kappa}(\mathbf{p}-\mathbf{q}) },
\end{align}
with $ q = (\omega,p)$ and the  $\widetilde{\partial}_t$ denotes the scale derivative acting only on the regulator as it is standard in the Wetterich formalism.
\begin{equation}
\label{macroshear}
\partial_t\eta_{\kappa} = \left. \frac{1}{2} \frac{\partial}{\partial p^{2}} \, P_{ij}^{T}(\mathbf{p}) \, \partial_t \Gamma^{(2)}_{\kappa,\bar{u}_i u_j}(0,\mathbf{p}) \right|_{\mathbf{p}=0}.
\end{equation} 
Comparing with \eqn{microshear} of the previous section and 
Matching symmetries, it is clear that \eqn{macroshear} can be thought of as the anomalous dissipation correction
to the shear viscosity. The \eqn{microshear} on the other hand should be thought as a microscopic statistical fluctuation (as it occurs in the compressible regime) corresponding to the quantum numbers of a shear.
\section{Average evolution via linear response}
The Ward identities described in the previous section prove one constraint relating the average and the fluctuation of the distribution.   The second constraint is the linear response equation
\begin{equation}
\label{lin2}
\ave{J_i(x,t+\Delta t)} = \int_{t}^{t+\Delta t} dx^\prime  dt^\prime \left[ \mathcal{D}_{ik}(x-x^\prime ,t-t^\prime )\ave{J_k(x^\prime ,t')}+\mathcal{M}_{ikl}(x-x^\prime ,t-t^\prime ) \ave{T^{kl}(x^\prime ,t^\prime )}\right]
\end{equation}
\begin{equation}\label{lin1}
\ave{T_{ik}(x,t+\Delta t)} = \int_{t}^{t+\Delta t} dx^\prime  dt^\prime \left[ \mathcal{C}_{iklm}(x-x^\prime ,t-t^\prime )\ave{T_{lm}(x^\prime ,t')}+\mathcal{M}^T_{ikm}(x-x^\prime ,t-t^\prime ) \ave{J^{m}(x^\prime ,t^\prime )}\right]
\end{equation}
where the $\mathcal{F}$ (representing $\mathcal{C},\mathcal{D},\mathcal{M}$) is, in analogous notation \cite{gaussian,gaussiandiff}, obtained from the fourier transforms of their corresponding $F$ (representing $C,M,D$ of \eqn{partition}) via \eqn{linrespdef}. Here we note that equations \eqn{lin1} and \eqn{lin2}
are convolutions in configuration space, so their Fourier transforms are multiplications
\begin{align}
\label{convlin}
\ave{\tilde{J}_i(k,t+\Delta t)} &= \int_{t}^{t+\Delta t}  dt^\prime \left[ \tilde{\mathcal{D}}_{ik}(k^\prime ,t-t^\prime ) \ave{\tilde{J}_k(k,t') } + \tilde{\mathcal{M}}_{ikl} (k,t-t^\prime ) \ave{\tilde{T}^{kl}(k,t^\prime )}\right] \\ 
\ave{\tilde{T}_{ik}(x,t+\Delta t)} &= \int_{t}^{t+\Delta t} dt^\prime \left[ \tilde{\mathcal{C}}_{iklm}(k,t-t^\prime )\ave{\tilde{T}^{lm}(k,t')} + \tilde{\mathcal{M}}^T_{ikm}(k,t-t^\prime ) \ave{\tilde{J}^{m}(k,t^\prime )}\right]    
\end{align}
And, as explained in the previous section, the correlation functions above and below $k_0$ are very different, with the microscopic ones being compressible and macroscopic ones being incompressible.

\begin{equation}
\ave{\tilde{J}_i(k,t)} =\left\{
\begin{array}{cc}
\rho_{0}\, u_{i}^{T} + \left\langle \Delta J_{i}^{\mathrm{comp} \rightarrow\mathrm{inc}}
\right\rangle  & k\leq k_0 \\
\rho_{0} \left( \delta_{ij} -
\frac{k_{i}k_{j}}{k^{2}} \right)
u_{j}(k,t), & k\geq k_0
\end{array}
\right. 
\end{equation}
where

\begin{equation}
\left\langle
\Delta J_{i}^{\mathrm{comp}\rightarrow\mathrm{inc}}(p) \right\rangle \simeq \lim_{\substack{|p|/k_{0}\rightarrow 0\\[2pt] |\omega|/(c_{s}k_{0})\rightarrow 0}} P_{ij}^{T}(p)
\int_{q\in\mathrm{comp}} V_{\kappa,jAB}(p,q,-p-q)\,
G_{\kappa,\mathrm{comp}}^{AB}(q)\, \Phi_{\mathrm{macro}}(p)
+ \mathcal{O}\!\left( \Phi_{\mathrm{macro}}^{2}
\right)
\end{equation}
where $V_{\kappa,jAB}$ is the one-particle-irreducible three-point of full compressible vertex\footnote{A,B run over all compressible microscopic degrees of freedom of the FRG field multiplet.}. This ``compressible current`` $J_{i}^{\mathrm{comp}\rightarrow\mathrm{inc}}$ arising from the counterterm can be thought of as the anomalous dissipation diffusive current, creating a turbulent flow of molecules from the chaotic microscopic molecular motion, at the scale, comparable with molecular distance, where dynamics is also compressible. The energy-momentum tensor is more complicated but follows the same pattern
\begin{equation}
\ave{\tilde{T}_{ki}(k,t)} =\left\{
\begin{array}{cc}
\langle \rho_{0} \int_{q} u_{k}^{T}(q,t)\,
u_{i}^{T}(k-q,t) - i\eta
\left( k_{k}u_{i}^{T} + k_{i}u_{k}^{T}
\right) - \\  \rho_{0}\, \delta_{ki}\,
k_{m}k_{n}/k^{2} \int_{q} u_{m}^{T}(q,t)\, u_{n}^{T}(k-q,t) \rangle
+ \left\langle \Delta T_{ki}^{\mathrm{comp} \rightarrow\mathrm{inc}}
\right\rangle   & k\leq k_0\\
\rho\,u_{k}u_{i}
+ p(\rho,s)\,\delta_{ki}
- \eta \left( \partial_{k}u_{i}
+ \partial_{i}u_{k}
- \frac{2}{3}\, \delta_{ki}\,
\partial_{\ell}u_{\ell}
\right) - \zeta\, \delta_{ki}\,
\partial_{\ell}u_{\ell}  & k\geq k_0
\end{array}
\right.
\end{equation}
The anomalously dissipative energy momentum tensor, therefore, should generally be expanded as
\begin{align}
& \left\langle \Delta T_{ki}^{\mathrm{comp} \rightarrow\mathrm{inc}}(p) \right\rangle \simeq \int_{q} V_{\kappa,ki;AB}^{(1)}(p,q,-p-q)\, G_{\kappa,\mathrm{comp} }^{AB}(q)\, \Phi_{\mathrm{macro}}(p) + \frac{1}{2}  \int_{p_{1}}
\int_{q\in\mathrm{comp}} \Phi_{\mathrm{macro},C}(p_{1}) \nonumber \\ 
& \times  G_{\kappa,\mathrm{comp}}^{AB}(q) V_{\kappa,ki;AB;CD}^{(2)}
\left( p,q,-q-p; p_{1},p-p_{1} \right) \,
\Phi_{\mathrm{macro},D}(p-p_{1}) + \mathcal{O}\!\left( \Phi_{\mathrm{macro}}^{3} \right)
\end{align}
where $V_{\kappa,ki;AB}^{(1)}(p,q,-p-q)$ connects one external incompressible stress insertion, one external macroscopic field, and
two internal compressible fields. The $G_{\kappa,\mathrm{comp}} V^{(2)}$ carries the memory of the eliminated compressible physics into the effective incompressible stress tensor. It measures how the eliminated compressible microscopic sector modifies the quadratic constitutive relation between two macroscopic incompressible fields. The first line modifies the stress terms linear in the macroscopic field, for example the effective shear response. The second line modifies the quadratic convective and pressure structures.
as well as the correlators

We now proceed to find the anomalous contribution to the fluctuations, which very roughly corresponds to ``the wildness`` (\cite{wild}),the backreaction of the macroscopic turbulence on microscopic fluctuations and in general microstate distributions.

\begin{equation}
\tilde{D}_{ijkl}(k,t) =\left\{
\begin{array}{cc}
P_{ik}^{T} \, D_{T}^{\mathrm{inc}} 
- D_{\kappa}^{\mathrm{res}}  & k\leq k_0\\
P_{ik}\, D_{T} + k_{i} k_{k}\, D_{L}  & k\geq k_0
\end{array}
\right. 
\end{equation}
where $D_{T}  = \frac{2T\, P^{T}_{ik} (\delta^{ij} - k^i k^j)  \eta\, } {\omega^{2}
+\left(\eta k^{2}/\rho \right)^{2}} $, and $ D_{L}  = \frac{ 2 T k^i k^j \omega^{2} \left[ \zeta + \frac{4}{3}\,\eta \right]  }
{\left(c_{s}^{2}k^{2}-\omega^{2}\right)^{2}
+\gamma_{L}^{2}\omega^{2}k^{4}}$.
\begin{equation}
\tilde{M}_{kim}(k,t) =\left\{
\begin{array}{cc}
M_{ij\mid k}^{\mathrm{inc}}  - 
M_{\kappa}^{\mathrm{res}} & k\leq k_0\\
\frac{\omega}{k^{2}} \left[ k_{i}P_{jk}^{T}
+ k_{j}P_{ik}^{T} \right] D_{T}  + H_{ij} \frac{k_{k}}{k} \, D_{L}   & k\geq k_0
\end{array} 
\right. 
\end{equation}
where $H_{ij} = \left[ \frac{\omega}{k^{2}} k_{i}k_{j} + r  P_{ij}^{T} \right]$ and $r  = \frac{3 \zeta + 4 \eta}
{3 \zeta - 2 \eta}\, \frac{k}{\omega}
+ \frac{ 3 \zeta + 4 \eta} {2\eta}\, \frac{\omega}{ 3 c_{s}^{2}k}$. Equivalently, defining the pressure-projected tensor $ Q_{ij\mid mn}(k) = \frac{1}{2} \left( \delta_{im}\delta_{jn} + \delta_{in}\delta_{jm} \right) - \delta_{ij}\, \frac{k_{m}k_{n}}{k^{2}}$, the nonlinear terms can be combined as
\begin{equation}
M_{ij\mid k}^{\mathrm{inc}}(p)
= -\frac{i\eta}{\rho_{0}} \left[ k_{i}\, D_{jk}^{\mathrm{inc}}(p) + k_{j}\, D_{ik}^{\mathrm{inc}}(p) \right] + \rho_{0}^{2}\, Q_{ij\mid mn}(k) \int_{q} \left\langle v_{m}(q)\, v_{n}(p-q)\, v_{k}(-p) \right\rangle_{c}.
\end{equation} 
At a centered Gaussian level, $ \left\langle
v_{m}\, v_{n}\, v_{k} \right\rangle_{c} = 0$, so only at that approximation,
\begin{equation}
M_{ij\mid k}^{\mathrm{inc},G}(p)
=
-\frac{i\eta}{\rho_{0}}
\left[
k_{i}\,
P_{jk}^{T}
+
k_{j}\,
P_{ik}^{T}
\right]
D_{T}^{\mathrm{inc}}(p).
\end{equation}
Beyond the Gaussian approximation, the convective and nonlocal-pressure terms are both controlled by the full connected three-velocity correlator.

\begin{equation}
\tilde{C}_{kim}(k,t) =\left\{
\begin{array}{cc}
C_{ijlk}  -  C_{\kappa,ijlk}^{\mathrm{res}} & k\leq k_0\\
\frac{\omega^{2}}{k^{4}}
D_{T} \left( k_{i}k_{k} P_{jl}^{T}
+ k_{i}k_{l}P_{jk}^{T} + k_{j}k_{k}P_{il}^{T} + k_{j}k_{l}P_{ik}^{T} \right) + k^{2} D_{L} H_{ij} H_{kl} + N_{ij\mid kl}^{\perp} & k\geq k_0
\end{array}
\right.
\end{equation}
where $N_{ij\mid kl}^{\perp} = 2T \left[
\eta \left( P_{ik}^{T}P_{jl}^{T} + P_{il}^{T}P_{jk}^{T} \right) + A_{2\eta\lambda}\,
P_{ij}^{T}P_{kl}^{T} \right]$ and $
T_{ij}[v](p) = -i\eta \left(
k_{i}v_{j} + k_{j}v_{i} \right) + \rho_{0} \int_{q} \left( \delta_{im}\delta_{jn}
- \delta_{ij} \frac{k_{m}k_{n}}{k^{2}}
\right) v_{m}(q)\, v_{n}(p-q)$.

\begin{equation}
\begin{pmatrix}
D_{\kappa}^{\mathrm{res}} &
M_{\kappa}^{\mathrm{res}\,T}
\\[2mm]
M_{\kappa}^{\mathrm{res}} &
C_{\kappa}^{\mathrm{res}}
\end{pmatrix} =
\begin{pmatrix}
D_{\kappa} &
M_{\kappa}^{T}
\\[2mm]
M_{\kappa} &
C_{\kappa}
\end{pmatrix}
\begin{pmatrix}
\Sigma_{\kappa}^{JJ} &
\Sigma_{\kappa}^{JT}
\\[2mm]
\Sigma_{\kappa}^{TJ} &
\Sigma_{\kappa}^{TT}
\end{pmatrix}
\begin{pmatrix}
D_{\kappa} &
M_{\kappa}^{T}
\\[2mm]
M_{\kappa} &
C_{\kappa}
\end{pmatrix}
+ \mathcal{O}(\Sigma^{2}).
\end{equation}
Compressible propagators therefore remain present in internal loops even though the external correlation functions carry only transverse hydrodynamic indices.

This makes the convolution integrals evolve slow and fast perturbations in very different ways. The microscopic compressible correlators do not appear as separate additive correlators in the infrared theory, but integrating them out generally changes the effective incompressible action and therefore changes the incompressible correlation functions indirectly. In general,
$ \{ D_{L}^{\mathrm{comp}},M_{ij\mid k}^{\mathrm{comp}}, C_{ij\mid kl}^{\mathrm{comp}} \} \neq 0 $. The density, longitudinal current and isotropic stress are coupled through the continuity equation and the equation of state. This longitudinal compressible sector is precisely what disappears as a propagating macroscopic degree of freedom below $k_{0}$. The residue must not be evaluated simply by putting the microscopic fields equal to zero. It is obtained by first retaining their coupling to the macroscopic sector and then taking the external low-energy limit:
\begin{equation}
\frac{|p|}{k_{0}}
\rightarrow
0,
\qquad
\frac{|\omega|}{c_{s}k_{0}}
\rightarrow
0.
\end{equation}
The internal microscopic fluctuations remain at momenta and frequencies characteristic of the compressible sector $ G_{\kappa}^{\mathrm{inc}}
\neq G_{\mathrm{NS}}
$ but it is also wrong to write $ G_{\kappa}^{\mathrm{inc}}
= G_{\mathrm{NS}}
+ G_{\mathrm{micro}}$. Integrating out the microscopic sector produces 
\begin{equation}
G_{\mathrm{eff},\kappa}^{\mathrm{inc}}
= \left[ G_{\kappa}^{\mathrm{inc}} -\Sigma_{\mathrm{comp} \rightarrow\mathrm{inc}} \right].
\end{equation}
The microscopic sector modifies the equation of motion rather than the observed fluctuations directly. The observed fluctuations are computed afterwards from the modified equation of motion. This is the mathematically precise place where the compressible microscopic sector manifests itself. 
\begin{equation}
\Sigma_{\mathrm{comp}\rightarrow\mathrm{inc}}
(p;k_{0}) = \lim_{\substack{|p|/k_{0}\rightarrow 0\\[2pt]
|\omega|/(c_{s}k_{0})\rightarrow 0}}
\int_{\mathrm{comp}}
V_{\mathrm{inc-comp}}\,
G_{\mathrm{comp}}(q)\,
G_{\mathrm{comp}}(q-p)\,
V_{\mathrm{comp-inc}}.
\end{equation}
The complete decoupling if $ \lim_{|p|/k_{0}\rightarrow 0}
\Sigma_{\kappa}^{\mathrm{comp}}(p)
= 0$ then the compressible sector leaves no infrared residue $ G_{\mathrm{eff, \kappa}}^{\mathrm{inc}} 
\rightarrow G_{\kappa}^{\mathrm{inc}}$. The convolution integrals treat slow incompressible perturbations and fast compressible perturbations very differently because they occupy different positions in the FRG loop. Instead of being governed by two unrelated RG evolutions, they participate in the same flow equation with different propagator blocks, different dispersion relations, and different momentum-frequency regions.

\section{Discussion and outlook}
While this work is a highly incomplete first effort, it lays out the ingredients required for an inherently generally stochastic hydrodynamics, implementing both Galilean symmetry and fluctuations non-perturbatively.   The  fundamental object is the Gaussian partition function \eqn{partition} evolves via the Galilean Ward identity, \cref{cac1,cac2,allward}.
Together with the linear response equation, \cref{lin2,lin1},  an initial ensemble of initial values for currents from which  $\ave{T_{ij}},\ave{J_{i}}, D_{ik}, M_{ijk}, C_{ijkl}$ can be calculated and a thermostatic partition function $\lnz(T,\mu)$ to Gaussian order (basically the energy and molecular density, heat capacity and susceptibility), one could evolve the whole ensemble in a way that respects Galilean symmetries, via a lattice algorithm shown in Fig 2,3 of \cite{gaussian}.

The non-relativistic limit is more complicated than \cite{gaussian} as necessitates a current and a chemical potential, which triples the number of correlators and doubles the number of currents, but the equations remain closed as before.

The  issue is that a Gaussian $\lnz(T,\mu)$ can not be defined for an incompressible fluid, since it would be non-analytic.  However this can be resolved by a functional renormalization group calculation, where macroscopic correlation functions look incompressible while hiding, in counterterms, the microscopic compressible structure.  Since linear response is non-local in frequency and wavenumber, such microscopic terms will generate fluctuations at the scale of the non-locality of pressure, cascading statistical fluctuations up to the macroscopic turbulence scale.   This has the potential of explaining spontaneous stochasticity, which would indeed be the molecular noise given by statistical mechanics amplified to macroscopic scales by a regime where the microscopic chaos of statistical mechanics and the macroscopic incompressible turbulence scale talk to each other. 

Incompressibility ensures that the volume preserving diffeomorphism invariance works in the same way in relativistic \cite{dubovsky} as in non-relativistic \cite{arnold} regimes:  It ensures local conservation of entropy, in that while dissipation and entropy creation of course  do occur, they can not be measured locally (on a scale set by the incompressibility scale) because it can not be locally distinguished from a fluctuation within a microstate distribution.
This ensures the turbulence scale and the the microscopic fluctuation scale, while different by orders of magnitude, are ultimately related, making spontaneous stochasticity natural.

Diffeomorphism invariance together with the inherently probabilistic nature of \eqn{partition} and Stochastic calculus gives a natural setting which includes spontaneous stochasticity \cite{spont1,spont2} automatically.   Anomalous dissipation \cite{andiss} (in this picture, irregular turbulent structures become features of a microstate distribution) and Wild/nightmare solutions \cite{wild} (in this picture, ''tails'' of statistical microstate distributions which become features of the turbulent flow) are then not mathematical abstractions but potentially observable phenomena to be incorporated into our understanding of statistical physics.   

In this regard, we note that the concept of ``weak solutions'' \cite{andiss,onsager} around which anomalous transport and Wild solutions \cite{wild} are defined, is somewhat analogous to the renormalization group (with the test function paralleling renormalization group coarse-graining, and notions of universality of coarse-graining vs universality w.r.t. test functions used the same way).
Thus, the renormalization group treatment of the compressibility issue has important consequences on the evolution of the average quantities $\ave{T_{ij}(x)},\ave{J_i(x)}$, which will survive any coarse-graining.
While distances of the order of the compressibility scale are unobservable (even in principle, as probing them generally destroys the fluid), their effect survives, via the residual, in macroscopic dynamics.   This should give visible effects, for it is expected spontaneous stochasticity is set around the compressibility scale, and the residual might be crucial in determining the interaction of this scale with the turbulence scale, as seen in \cite{spont2}. 

The  breaking of incompressibility at the  UV cutoff means that the counterterm should be treated as an anomaly (breaking the volume-preserving diffeomorphisms of \cite{arnold}), which might  alter the fluctuation distribution worked out via topology from macroscopic statistical mechanics alone \cite{braunstein2}, something familiar within quantum field theory\cite{peskin}. The incorporation of anomalies in the topological derivation of \cite{braunstein2} is therefore a promising further investigation.

It remains to be seen whether such an approach, and in general a non-perturbative treatment of fluctuations,is really needed to understand experimental data.    Calculating anomalous diffusion from first principles would be a possible but long-winded project, necessitating inputs from both hydrodynamics and microscopic statistical mechanics.  Hence, a more qualitative experimental signature for such dynamics would be useful.

The experimental-driven question which prompted \cite{gaussian} is the question of the existence of what appears to be a fluid with very few degrees of freedom \cite{zajc}.   In parallel, it was found that systems with very few strongly interacting ultracold atoms appear to be surprisingly fluid-like \cite{giacalone} \footnote{Note that it is a fluid rather than a super-fluid, since no evidence was found for energy gaps of fluidic excitations}, and even every-day systems with a surprisingly small number of particles behave as a fluid \cite{braznut}.   It is clear that concepts such as spontaneous stochasticity are related to the experimentally investigable question of ''what is the smallest fluid''.

If turbulence and spontaneous stochasticity are found in such small systems, their scales will be much closer to each other than in ordinary turbulent fluids.   Perturbative fluctuating hydrodynamics will fail by a simple order of magnitude estimate, but our approach might not.  From there, one might check if the corrections from inserting an inherent stochasticity into the fluids evolution rather than treating it as a perturbation on a deterministic equation is worthwhile, also for more fundamental and mathematical questions such as Onsager's conjecture and the existence of a low viscosity limit from anomalous dissipation.

This brings us back to the ''philosophical'' issues discussed in the introduction.   Some of the deepest questions usually associated with fluid dynamics are thought to lie within the realm of mathematics, but fluids are very much physical objects, depending on such ingredients as statistical mechanics for their definition.
A proper treatment of how statistical mechanics and fluid dynamics meet when viscosity is so low that backreaction to thermal fluctuations is non-negligible might be necessary to better understand phenomena such as turbulence and anomalous dissipation, also from a mathematical point of view.

\textbf{Acknowledgements} GT thanks Bolsa de produtividade CNPQ 305731/2023-8 and FAPESP 2023/06278-2 as well as participation in the tematico 2023/13749-1 for support.


\section*{Appendix}

\appendix

\section{Functional construction of the incompressible effective theory}\label{appendix1}

This appendix develops the functional construction used to connect the compressible microscopic description with the effective incompressible theory. The density and longitudinal-velocity
fluctuations are integrated out and their effect consequently remains encoded in the correlation functions of incompressible sector
\begin{equation}
Z_{\kappa}^{\mathrm{comp}}
[J,T] = \int
\mathcal D\delta\rho\, \mathcal D v^{L}\,
\mathcal D v^{T}\, \exp\!\left[
- S_{\kappa}^{\mathrm{comp}}
[\delta\rho,v^{L},v^{T}] + J_\rho\delta\rho + J_Lv^{L} + J_Tv^{T}
\right].
\label{eq:Zcomp}
\end{equation}
The transverse field is kept as an explicit infrared variable. It only means that the infrared observer does not probe them directly.
Their contribution is absorbed into an effective generating functional
for the transverse velocity by integrating over the compressible sector,
\begin{equation}
Z_{\kappa}^{\mathrm{inc}}[J_T] = \int \mathcal D\delta\rho\, \mathcal D v^{L}\, Z_{\kappa}^{\mathrm{comp}}
[J_\rho=0,J_L=0,J_T].
\label{eq:Zinc}
\end{equation}
Equation~\eqref{eq:Zinc} defines the incompressible theory as an effective theory. The boundary conditions in the notation of \eqn{genward} and using the renormalization group prescription of \eqn{renfourier} are
\begin{align}
& \frac{\partial \lnz}{\partial k_0} ={}
\Theta(k_0-|\mathbf{k}|) \Theta\! \left(\frac{k_0}{c} - |\omega| \right)
\frac{\partial M}{\partial k_0} +
\Theta(|\mathbf{k}|-k_0) \Theta\! \left(\frac{k_0}{c} - |\omega| \right) \frac{\partial N}{\partial k_0} +
\Theta\!\left(|\omega|-\frac{k_0}{c}\right)
\frac{\partial W}{\partial k_0} 
\nonumber\\
& + \delta(k_0-|\mathbf{k}|) \Theta\!\left(\frac{k_0}{c}-|\omega|\right) (M-N) + \frac{1}{c}
\delta\!\left(\frac{k_0}{c}-|\omega|\right) \left[ \Theta(k_0-|\mathbf{k}|)M + \Theta(|\mathbf{k}|-k_0)N - W  \right].
\end{align}
The region where the descriptions meeting are at $ |\mathbf{k}|=k_0,$ one requires $ M(\omega,k_0)
= N(\omega,k_0). $ At $ |\omega|=\frac{k_0}{c},$ one requires $ N\!\left(k,\frac{k_0}{c}\right) = W\!\left(k,\frac{k_0}{c}\right) $. However, the derivatives need not match: $ \left. \frac{\partial M}{\partial |\mathbf{k}|} \right|_{k_0^-} \neq \left.
\frac{\partial N}{\partial |\mathbf{k}|} \right|_{k_0^+}$ since their discontinuity  reflects a  nonanalytic change in its scale derivative.

The boundary conditions of \eqref{wett}
\begin{equation}\label{boundaryt}
\Gamma_{\kappa} = \Theta(k_0-|\mathbf{p}|)
\Theta\!\left(\frac{k_0}{c}-|\omega|\right)
\Gamma_{\kappa,\mathrm{inc}} + \Theta(|\mathbf{p}|-k_0)
\Theta\!\left(\frac{k_0}{c}-|\omega|\right)
\Gamma_{\kappa,\mathrm{comp}} +
\Theta\!\left(|\omega|-\frac{k_0}{c}\right)
\Gamma_{\kappa,E}.
\end{equation}
It must not be interpreted as a sum of three independent theories. The same scale-dependent functional is evaluated with different active degrees of freedom in the corresponding momentum-frequency domains. The effective actions are matched continuously at the crossover scale, $ \Gamma_{\kappa}^{\mathrm{inc}} \Big|_{k_0^-} = \Gamma_{\kappa}^{\mathrm{comp}} \Big|_{k_0^+}$. The discontinuity $ \partial_{\kappa} \Gamma_{\kappa}^{\mathrm{inc}} \Big|_{k_0^-} \neq \partial_{\kappa} \Gamma_{\kappa}^{\mathrm{comp}} \Big|_{k_0^+}$ is because the sets of active degrees of freedom are different.

\subsection{Compressible}

The continuity part constructed with scale-dependent coefficients
\begin{align}
& \Gamma_{\kappa}^{\mathrm{continuity}}
+ \Gamma_{\kappa}^{\mathrm{momentum}} = \int \bigg\{ \bar{\rho} [ Z_{\rho,\kappa}\,\partial_t \rho + \lambda_{\rho,\kappa}\,
\partial_i(\rho u_i) ] + 
\bar{u}_i [ Z_{u,\kappa}\,
\rho ( \partial_t u_i
+ \lambda_{u,\kappa}\,
u_j\partial_j u_i ) \nonumber \\
& + \partial_i p_{\kappa}(\rho) - \eta_{\kappa}\nabla^2u_i - ( \zeta_{\kappa} + 3 \eta_{\kappa})  \partial_i\partial_j u_j ] \bigg\}.
\end{align}
 The coefficients are kept scale dependent $\kappa$
because the effective equations change as fluctuations are integrated out. The $\Gamma_{\kappa}^{\mathrm{continuity}}$ and $\Gamma_{\kappa}^{\mathrm{momentum}}$ enforce local mass conservation and momentum balance, respectively. We expand around a homogeneous configuration $\rho=\rho_{0,\kappa} +\delta\rho$ and $u_i=0$. The quadratic continuity contribution is
\begin{align}
& \Gamma_{\kappa,\mathrm{continuity}}^{(2)} + \Gamma_{\kappa,\mathrm{momentum}}^{(2)} = \int_q \bar{\rho}(-q) \left[ -i Z_{\rho,\kappa}\,\omega\,\delta\rho(q)
+i\lambda_{\rho,\kappa}\,\rho_{0,\kappa}\, p_i u_i(q) \right] 
\nonumber \\ & \bar{u}(-q) [-i Z_{u,\kappa} \,\rho_{0,\kappa}\,
\omega\,u_i(q) +i c_{s,\kappa}^{\,2}\,
p_i\,\delta\rho(q) +\eta_\kappa p^2 u_i(q)
+ \left[ \zeta_\kappa
+ \frac{d-2}{d}\eta_\kappa
\right] p_i p_j u_j(q)],
\end{align}
where the coefficients \(Z_{\rho,\kappa}\), \(Z_{c,\kappa}\), and \(Z_{g,\kappa}\) are initially retained so that the renormalization of the density, velocity, and response fields can be tracked independently throughout the renormalization-group flow. The stochastic force, represented by

\begin{equation}
\Gamma_{\kappa}^{(2)} = - \int_q \Phi^>_a(-q)\,
P_{\kappa,ab}(q)\, \Phi^>_b(q), \qquad P_{\kappa}(q)
= \begin{pmatrix}
-i Z_{\rho,\kappa}\,\omega
& i c_{s,\kappa}^{\,2}\,p_i
\\[6pt]
i Z_{c,\kappa}\,p_j
& A_{\kappa,ij}(q)
\end{pmatrix}
\end{equation}
specifies the covariance driven
of the microscopic fluctuations that drive, but not
replace the deterministic compressible dynamics. The matrix $P_{\kappa}$ is the quadratic response operator for the coupled compressible variables. The $ A_{\kappa,ij}(q) = \left( -i Z_{g,\kappa}\omega + \rho_{0} \eta_{\kappa}\,p^{2} \right) \delta_{ij} + \rho_{0} \left( \zeta_{\kappa} + \frac{3}{2}\eta_{\kappa} \right) p_{i}p_{j}$. 
The regulator must preserve the response-field structure 
\begin{equation}
R_{\kappa,ij}^{g}(\mathbf{p})
= \frac{\kappa^2-p^2}{\rho_0} \, \Theta(\kappa^2-p^2) \left[
\eta_{\kappa}\delta_{ij} + \left( \zeta_{\kappa} + \frac{1}{3}\eta_{\kappa} \right) \frac{p_ip_j}{p^2} \right]
\end{equation}
and is responsible to define which fluctuations are allowed to participate in the effective theory at a given RG scale. In other words, this regulator suppresses the low-momentum response modes without changing their transverse and longitudinal tensor decomposition. The density fluctuations are coupled to the regulated momentum fluctuations through the continuity equation. The green function from \eqref{greenf} is  

\begin{equation}
G_{\kappa}(q)
=
\begin{pmatrix}
2\,P_{\kappa}^{-1}(q)\,
N_{\kappa}(q)\,
P_{\kappa}^{-T}(q)
&
P_{\kappa}^{-T}(q)
\\[8pt]
P_{\kappa}^{-1}(q)
&
0
\end{pmatrix}.
\end{equation}
The exact flow equation is
\begin{equation}
\partial_t \Gamma_{\kappa}
=
\frac{1}{2}\,
\operatorname{Tr}
\left[
G_{\kappa}\,
\partial_t R_{\kappa}
\right],
\qquad
t=\ln\!\left(\frac{\kappa}{\Lambda}\right).
\end{equation}
The trace contains the full regularized propagator. Functional derivatives of this equation generate the flow equations for the scale-dependent
vertices. Taking two functional derivatives of the Wetterich equation yields
\begin{equation}
\partial_t \Gamma_{\kappa,AB}^{(2)}(p)
= \operatorname{Tr} \!\left[
G_{\kappa}\, \Gamma_{\kappa,A}^{(3)} G_{\kappa}\, \Gamma_{\kappa,B}^{(3)} G_{\kappa}\, \partial_tR_{\kappa}
\right] -  \frac{1}{2} \operatorname{Tr}
\!\left[ G_{\kappa}\, \Gamma_{\kappa,AB}^{(4)}
G_{\kappa}\, \partial_tR_{\kappa} \right].
\label{eq:exact_two_point_flow}
\end{equation}
he first contribution contains two three-point vertices, while the second
contains one four-point vertex.

\subsection{Incompressible}

The incompressible theory is defined on the transverse velocity sector.

\begin{equation}
\Gamma_{\kappa,2} = \int_q \bar{u}_i(-q) P_{ij}^{T}(\mathbf{q}) \left[ -iZ_{\kappa}\omega + \nu_{\kappa}p^2 \right] u_j(q) - D_{\kappa}(\mathbf{q}) \bar{u}_i(-q) P_{ij}^{T}(\mathbf{q}) \bar{u}_j(q), 
\end{equation}
The coefficient $Z_{\kappa}$ controls the temporal response, while $\nu_{\kappa}$ is the running kinematic viscosity. The kernel
$D_{\kappa}(\mathbf p)$ determines the covariance of the transverse stochastic force. The nonlinear Navier–Stokes interaction is

\begin{equation}
\Gamma_{\kappa,3} = \frac12 \int_{q_1}\int_{q_2}\int_{q_3}
(2\pi)^{d+1} \delta(q_1+q_2+q_3) \, \bar{u}_i(q_1) V_{\kappa,ijl}(\mathbf{q}_1) u_j(q_2) u_l(q_3),
\end{equation}
where $ V_{\kappa,ijl}(\mathbf{p}) = \frac{i\lambda_{\kappa}}{2} \left[ p_jP_{il}^{T}(\mathbf{p}) + p_lP_{ij}^{T}(\mathbf{p}) \right] $ determines the nonlinear mode coupling and it is the incompressible Navier--Stokes cubic vertex, which removes the longitudinal component generated by the quadratic velocity product. The transverse projector contained in the vertex is the explicit remnant of the eliminated pressure. At zero background field, the Hessian
\begin{equation}
\Gamma_{\kappa}^{(2)}(q) = P_{ij}^{T}(\mathbf{p})
\begin{pmatrix}
0
&
-iZ_{\kappa}\omega+\nu_{\kappa}p^2
\\[4pt]
iZ_{\kappa}\omega+\nu_{\kappa}p^2
&
-2D_{\kappa}(\mathbf{p})
\end{pmatrix}.
\end{equation}
The regulator of \eqref{reg} acts only in the response sector and generates a transverse external propagator in the incompressible sector

\begin{equation}
\partial_tR_{\kappa}  = \nu_{\kappa} \left[ (2-\eta_{\nu}) \kappa^2 + \eta_{\nu}p^2 \right] \Theta(\kappa^2-p^2) 
\end{equation}
Define the viscosity anomalous dimension by $ \eta_{\nu} =
-\partial_t\ln\nu_{\kappa}, \ t=\ln\kappa$. The green function from \eqref{greenf} is 
\begin{equation}
G_{\kappa,ij}(q) = P_{ij}^{T}(\mathbf{p})
\begin{pmatrix}
\dfrac{2D_{\kappa}(\mathbf{p})}
{Z_{\kappa}^{2}\omega^{2}+A_{\kappa}^{2}(\mathbf{p})}
&
\dfrac{1}
{-iZ_{\kappa}\omega+A_{\kappa}(\mathbf{p})}
\\[12pt]
\dfrac{1}
{iZ_{\kappa}\omega+A_{\kappa}(\mathbf{p})}
&
0
\end{pmatrix}.
\end{equation}
The current covariance is $ D^{\mathrm{macro}}_{ij}(q) = \rho_0^2 G^{uu}_{\kappa,ij}(q)$. We start from \eqref{trace}
\begin{equation}\label{FRGin}
\partial_t\Gamma_{\kappa}[\Phi] = \frac{1}{2} \operatorname{Tr} \left[ G_{\kappa}[\Phi] \,\partial_tR_{\kappa} \right] =  \int_q \frac{A_{\kappa}(p)\, \partial_tR_{\kappa}(p)}{Z_{\kappa}^{2}\omega^{2} + A_{\kappa}^{2}(p)} = \frac{\nu_{\kappa}  \kappa^{5} }{3\pi Z_{\kappa}} \,,
\end{equation}
The \eqref{FRGin} is the flow of the effective average action evaluated at vanishing field, i.e., the vacuum contribution to the Wetterich equation. It is the simplest projection of the FRG flow. this expression alone cannot produce the running of the couplings. The first nontrivial physical flow is the flow of the two-point vertex,
\begin{align}
\partial_t \Gamma_{\kappa}^{(2)}
&= -\frac12 \operatorname{Tr}
\left[ G_{\kappa} \left( \Gamma_{\kappa}^{(4)}
- 2 \Gamma_{\kappa}^{(3)} G_{\kappa} \Gamma_{\kappa}^{(3)} \right) G_{\kappa}
\, \partial_tR_{\kappa} \right]. \\
& = \text{two-cubic-vertex diagram}
- \frac{1}{2}\, \text{quartic-vertex diagram}.
\end{align}
The effective actions are matched continuously at the crossover scale, $ \Gamma_{\kappa}^{\mathrm{inc}} \Big|_{k_0^-} = \Gamma_{\kappa}^{\mathrm{comp}} \Big|_{k_0^+}$ the continuity condition guarantees that the same observable physics is described on the matching surface. $ \partial_{\kappa}
\Gamma_{\kappa}^{\mathrm{inc}} \Big|_{k_0^-} \neq \partial_{\kappa} \Gamma_{\kappa}^{\mathrm{comp}}
\Big|_{k_0^+}$. This discontinuity of the scale derivative records the abrupt change in the set of active degrees of freedom. It is not enough to determine the residue without the macro-micro coupling. 
\begin{equation}
\partial_t \Gamma_{\kappa}^{(2)}
= \operatorname{Tr} \left[ G_{\kappa} \Gamma_{\kappa}^{(3)} G_{\kappa} \Gamma_{\kappa}^{(3)} G_{\kappa} \, \partial_tR_{\kappa} \right].
\end{equation}
This is the two-cubic-vertex contribution to the incompressible two-point flow and It should not be identified directly with the compressible residue.

\section{Compressible residue kernels}

Then,
\begin{align}
\Sigma_{\kappa,ik}^{JJ}(p) = 
\int_{q\in\mathcal{C}(k_0)} \Big[
& \partial_t G_{\kappa,\rho\rho}^{>}(q)\, G_{\kappa,ik}^{>}(r)
+ G_{\kappa,\rho k}^{>}(q)\, \partial_t G_{\kappa,i\rho}^{>}(r) + G_{\kappa,i\rho}^{>}(q)\,
\partial_t G_{\kappa,\rho k}^{>}(r) \ +
\nonumber \\ &
 G_{\kappa,ik}^{>}(q)\,
\partial_t G_{\kappa,\rho\rho}^{>}(r)
\Big], \qquad r=p-q.
\end{align}

Then,
\begin{equation}
\Sigma_{\kappa,i\mid kl}^{JT}(p;k_0)
= \partial_t \int_{q\in\mathcal{C}(k_0)}
V_{\kappa,J_i;ab} \bigl(p;q,p-q\bigr)\,
G_{\kappa,ac}^{>}(q)\, G_{\kappa,bd}^{>}(p-q)\,
V_{\kappa,T_{kl};cd} \bigl(-p;q-p,-q\bigr).
\end{equation}
This is not generally an independent kernel. For a real Gaussian theory with the usual exchange properties,
$ \Sigma_{\kappa,ij\mid k}^{TJ}(p) =
\Sigma_{\kappa,k\mid ij}^{JT}(-p)$. In matrix notation $ \Sigma_{\kappa}^{TJ}(p) = \left[ \Sigma_{\kappa}^{JT}(-p) \right]^{T}$. Then,
\begin{equation}
\Sigma_{\kappa,ij\mid k}^{TJ}(p;k_0)
=
\partial_t
\int_{q\in\mathcal{C}(k_0)}
V_{\kappa,T_{ij};ab}
\bigl(p;q,p-q\bigr)\,
G_{\kappa,ac}^{>}(q)\,
G_{\kappa,bd}^{>}(p-q)\,
V_{\kappa,J_k;cd}
\bigl(-p;q-p,-q\bigr).
\end{equation}

\begin{align}
\Sigma_{\kappa,ij\mid kl}^{TT}(p;k_0)
& = \lim_{\substack{|p|/k_0\rightarrow 0\\[2pt]
|\omega|/(c_s k_0)\rightarrow 0}}
\,  \int_{q\in\mathcal{C}(k_0)}
\partial \cdot V_{\kappa,T_{ij} ;AB}^{\mathrm{inc\text{-}comp}}
\bigl(p;q,p-q\bigr)\,
G_{\kappa,AC}^{\mathrm{>}}(q)\,
G_{\kappa,BD}^{\mathrm{<}}(p-q) \times \nonumber \\ &
V_{\kappa,T_{kl};CD}^{\mathrm{comp\text{-}inc}}
\bigl(-p;q-p,-q\bigr) + 
G_{\kappa,AC}^{\mathrm{<}}(q)\, \partial \cdot V_{\kappa,T_{ij};AB}^{\mathrm{inc\text{-}comp}}
\bigl(p;q,p-q\bigr)\
G_{\kappa,BD}^{\mathrm{<}}(p-q)
\end{align}


\end{document}